\documentclass{ifacconf}

\usepackage{algorithm}
\usepackage[noend]{algpseudocode}
\usepackage{amsmath}
\usepackage{amsfonts}
\usepackage{graphicx}
\usepackage{url}
\usepackage{amssymb}
\usepackage{tikz}
\usepackage{mathdots}
\usepackage{cancel}
\usepackage{color}
\usepackage{siunitx}
\usepackage{array}
\usepackage{subcaption}
\usepackage{multirow}
\usepackage{tabularx}
\usepackage{booktabs}
\usepackage{colortbl}
\usepackage{arydshln} 
\usepackage{changepage} 
\usepackage{wrapfig}
 
\newcommand{\ie}{\emph{i.e., }}
\newcommand{\eg}{\emph{e.g., }}

\newtheorem{remark}{Remark}
\newtheorem{definition}{Definition}

\newtheorem{theorem}{Theorem}[section]
\newtheorem{corollary}{Corollary}[section]

\usepackage{graphicx}      
\usepackage{natbib}        
\begin{document}
\begin{frontmatter}

\title{Safe Reach Set Computation\\
via Neural Barrier Certificates\thanksref{footnoteinfo}}

\thanks[footnoteinfo]{Sergiy Bogomolov, contributing author, deceased in 2023.}

\author[First]{Alessandro Abate} 
\author[Second]{Sergiy Bogomolov} 
\author[First]{Alec Edwards}
\author[Second]{Kostiantyn Potomkin}
\author[Third]{Sadegh Soudjani}
\author[Fourth]{Paolo Zuliani}

\address[First]{University of Oxford, Oxford, UK}
\address[Second]{Newcastle University, Newcastle, UK}
\address[Third]{Max Planck Institute for Software Systems, Germany}
\address[Fourth]{Universit\`a di Roma ``La Sapienza'',
Rome, Italy}

\begin{abstract}          
We present a novel technique for online safety verification of autonomous systems, which performs reachability analysis efficiently for both bounded and unbounded horizons by employing neural barrier certificates. Our approach uses barrier certificates given by parameterized neural networks that depend on a given initial set, unsafe sets, and time horizon.
Such networks are trained efficiently offline using system simulations sampled from regions of the state space. We then employ a meta-neural network to generalize the barrier certificates to state-space regions that are outside the training set. These certificates are generated and validated online as sound over-approximations of the reachable states, thus either ensuring system safety or activating appropriate alternative actions in unsafe scenarios. We demonstrate our technique on case studies from linear models to nonlinear control-dependent models for online autonomous driving scenarios. 
\end{abstract}

\begin{keyword}
reachability, safety, soundness, barrier certificates, neural networks, verification, control, online autonomous systems
\end{keyword}

\end{frontmatter}

\section{Introduction}
Reachability analysis, \ie determining the regions of the state space the temporal dynamics of a system eventually enters, is crucial for establishing safety of a system. Unfortunately, this problem is in general undecidable except for a restricted class of linear dynamical models \citep{Alur1993,HENZINGER1998}. Furthermore, algorithmic solutions for sound relaxations of the problem \citep{Franzle19} typically suffer from either the ``curse of dimensionality'' (exceeding time/space complexity with respect to system size) or limited practical utility (very coarse over-approximations returning too many ``false alarms''). One reason is because many reachability techniques require the explicit computation of the system dynamics, which most often entails integrating ordinary differential equations (ODEs). Symbolic integration is restricted to limited  cases of ODEs, and in general numerical integration is hard  -- solving Lipschitz continuous ODEs over compact domains is a PSPACE-complete problem \citep{Kawamura09}.

Inductive methods for reachability offer an alternative approach in which one looks for a {\em barrier certificate}
\citep{Tiwari03,BCPrajnaJ04}, a function of the system state whose zero-level set separates the ``unsafe'' region from the system trajectories that start from a given initial set. The existence of a barrier certificate entails that the system is safe. 
Three advantages of barrier certificates are that they can establish safety: 1) over infinite time horizons, 2) for systems with nonlinear dynamics, and 3) uncertain inputs or parameters. Reachability methods that offer some or all of these advantages do exist, but are usually  computationally expensive.
In contrast, checking that a candidate function is indeed a barrier certificate can be done in many cases efficiently and automatically. However, finding barrier certificates is much harder, and this is a disadvantage of the approach. 

In this work, we provide a novel approach to synthesize, via barrier certificates, sound over-approximations of reach sets for dynamical systems, whether linear or non-linear, possibly under control inputs. We go beyond establishing safety by computing valid over-approximations {\em without} explicitly solving the system dynamics.
We devise a two-level technique which utilizes neural networks as efficient emulators of both the system dynamics and barrier certificates. Since barrier certificate candidates are efficiently generated via neural networks, our approach enables online computation of reach sets after appropriate offline training. Key to our approach is a meta-neural network {\em MetaNN} that generalizes barrier certificates over varying initial and unsafe sets.
We use FOSSIL~\citep{fossil21} as the  underlying generator of example barrier certificates with correctness guarantees. We note that FOSSIL does not scale to online use due to the intrinsic complexity of the algorithms implemented -- this motivates the introduction of our MetaNN. The overall soundness of our approach is guaranteed by utilizing Satisfiability Modulo Theories (SMT) solvers for checking candidate barrier certificates.



We show that our approach successfully generates sound over-approximations of reach sets of linear and non-linear systems. The efficiency of reach-sets generation is demonstrated in an online autonomous driving scenario. Our approach not only extends the capabilities of barrier certificate verification to the synthesis of parameterized controllers, but also increases the success rate of online barrier certificate generation from $78\%$ to $99\%$ compared to direct application of FOSSIL (at a cost of 10 hours of offline training on a standard laptop). 




To summarize, the contributions of this paper are:
\begin{itemize}
    \item We show that level sets of barrier certificates can be used to over-approximate reach sets. 
    \item We consider barrier certificates as functions of initial and unsafe sets and use them for reach set computations.
    \item We provide a computational framework that fixes parameterized templates for initial and unsafe sets and computes a MetaNN as generator of barrier certificates.
    \item We employ offline training and fast online validation and execution that increase the success rate of generating barrier certificates for safe online autonomous driving. 
\end{itemize}






\noindent\textbf{Related Work.}
Verifying safe behavior of autonomous systems is an important, timely topic which is receiving much attention from the research community \citep{Fisher21}. In particular, one area of interest include the synthesis of provably safe controllers for autonomous systems (\eg \citep{FanCAV20,Johnson2016}) possibly featuring machine learning components, such as controllers based on neural networks \citep{Radoslav20,Radoslav21}, reinforcement learning \citep{CPSReLe19,kazemi2020formal}, data-driven model predictive control \citep{Rosolia18}, or subject to perceptual uncertainty \citep{JhaRSS18}.

Barrier certificates, first introduced by
\cite{prajna2006BarrierCertificatesNonlinear}, can be used to certify safety of dynamical systems. The computations
	rely on convex optimization and hence can result in solutions which are numerically sensitive and unsound. 
	A recent approach aims at synthesizing weaker barrier certificates using bilinear matrix inequalities \citep{WangCAV21}, while the work by \cite{AmesXGT17}  seeks to include performance requirements in barrier certificates.
%
 Usage of counter-examples within a counter-example guided inductive synthesis (CEGIS) framework is a powerful and common approach for synthesizing barrier certificates \citep{peruffo2021automated,fossil21,heersink2021FormalVerificationOctorotor,ratschan2018SimulationBasedComputation,Samari24}.
Learning barrier certificates using data is also studied in \citep{nejati2023formal} for deterministic systems and in \citep{schon2024data} for stochastic systems.
	Neural networks are utilized to synthesize barrier certificates, \eg in the works by
\cite{jin2020NeuralCertificatesSafe} and \cite{zhao2020SynthesizingBarrierCertificates,zhao2021LearningSafeNeural,zhao2021SynthesizingReLUNeural}. However, these approaches are either unsound \citep{jin2020NeuralCertificatesSafe}, or limited to the usage of ReLU-based activation functions.
The paper by \cite{CBFml20} is close in spirit to our approach, although it uses support vector machines and is limited to a specific class of systems (affine control robotic systems), while we do not pose such restrictions.
	Our proposed framework may leverage any tool that generates certified barrier certificates. We employ FOSSIL \citep{fossil21} due to its ability to provide \textit{training data} (\ie example barrier certificates) with correctness guarantees.

 %
%



\section{Preliminaries} \label{sec:prelim}

\noindent\textbf{Feed-Forward Neural Networks.}
Consider a network $\mathcal{N}$ with $h_0$ input neurons, $k$ hidden layers each with $h_1, \ldots, h_k$ neurons and an output layer of $h_{k+1}$ neurons. We consider each layer to be
\noindent fully connected and denote the corresponding weight matrix for the $i$-th layer as $W_i$ and bias vector as $b_i$ for $i = 0, \ldots, k+1$, where $i=0$ denotes the input layer and $i=k+1$ the output layer. 
Each hidden layer $i$ has a corresponding activation function $\sigma_i : \mathbb{R} \to \mathbb{R}$ 
such that the output of the $i$-th layer is given by $z_i = \sigma_i (W_i z_{i-1} + b_i)$.




\noindent\textbf{Barrier Certificates.}
We are interested in the safety verification of autonomous systems. Consider a dynamical system modelled by
\begin{equation}
\label{eq:system}
    \frac{d}{dt}x(t) = f(x(t)), \quad t\in\mathbb R_{\ge 0},
\end{equation}
with system state $x \in \mathcal X\subseteq \mathbb R^n$ and vector field $f:\mathcal X\rightarrow \mathbb R^n$. The differential equation~\eqref{eq:system} has a unique, continuous solution from any initial state $x(0) = x_0\in\mathcal X$ under standard Lipschitz continuity of the vector field $f(\cdot)$. 

\begin{definition}\label{def:safety}
Given an initial set $\mathcal X_0\subseteq \mathcal X$ and an unsafe set $\mathcal X_u\subseteq \mathcal X$, the system \eqref{eq:system} is called {\em safe} if all trajectories starting from any initial state $x_0\in \mathcal X_0$ do not enter the unsafe set $\mathcal X_u$.
\end{definition}

\begin{definition}\label{def:BC}
Given an initial set $\mathcal X_0$ and unsafe set $\mathcal X_u$, a function $B:\mathcal X\rightarrow\mathbb R$ is called a {\em barrier certificate} for the system \eqref{eq:system} if it satisfies
\begin{align}\label{eq:barrier}
    & B(x) \leq 0 \ \forall x  \in \mathcal X_0 \text{ and }
    B(x) > 0 \ \forall x  \in \mathcal X_u, \\
    & \frac{\partial B(x)}{\partial x}f(x) < 0, \ \forall x \in \mathcal X \ s.t. \ B(x) = 0.\nonumber
\end{align}
\end{definition}

\begin{theorem}\citep{BCPrajnaJ04,prajna2007FrameworkWorstCaseStochastic}
\label{thm:BS}
Given the initial set $\mathcal X_0$ and unsafe set $\mathcal X_u$, the system \eqref{eq:system} is safe if there exists a barrier certificate $B(\cdot)$ satisfying the conditions \eqref{eq:barrier}.
\end{theorem}
The characterization of a barrier certificate $B$ relies on its \textit{Lie derivative}, which appears in the last requirement of \eqref{eq:barrier}. The Lie derivative of a continuously differentiable scalar function $B$ with respect to a vector field $f$ is 
\begin{equation}\label{eq-lie_der}
    \dot{B}(x) = \nabla B(x) \cdot f(x) = \sum_{i=1}^{n}\frac{\partial B}{\partial x_i} f_i(x),
\end{equation}
where $x_i$ is the $i^{\text{th}}$ element of the state: $x = (x_1,x_2,\ldots,x_n)$.
The Lie derivative describes how the value of the barrier certificate varies along the flow of $f$, namely along trajectories of the model.
Indeed, if we consider a trajectory $x(t)$, with $x(0) \in \mathcal{X}_0$, and the value of $B(x)$ along this trajectory, the first condition ensures $B(x(0))\le 0$.
The condition on the Lie derivative then guarantees that the evolution of $x(t)$ cannot make $B(x)$ become positive, and hence $x(t)$ cannot enter $\mathcal{X}_U$ (where $B(x)$ is positive).


\noindent\textbf{Sound Synthesis of Barrier Certificates.}
Our proposed framework is applicable to any technique that generates sound barrier certificates. We employ the FOSSIL tool \citep{fossil21}, which uses a CEGIS loop 
that consists of two complementary components: the learner and the verifier. 
The learner is a neural network, of which the structure and characteristics (\ie layers and activation functions) can be selected freely within the tool.
 This flexibility allows for the construction of reach sets of varying structural complexity.
%
The verifier is an SMT-solver that can check the satisfiability of formulae over the real numbers. It provides correctness guarantees of the candidate barrier certificates proposed by the learner, or alternatively returns counter-examples to the learner: these are points within the region of interest where the candidate barrier certificate is invalid. This is achieved by checking the satisfiability of the negation of the conditions in \eqref{eq:barrier}. FOSSIL leverages two choices of SMT-solvers: Z3 \citep{demoura2008Z3EfficientSMT}, which is limited to reasoning over polynomial functions, and dReal \citep{gao2013DRealSMTSolver}, which can handle general nonlinear functions. 

\section{Barrier Certificates for Reach Sets}\label{bc4rs}

Barrier certificates demonstrate the existence of an invariant set in the vector field $f$, via the final condition of \eqref{eq:barrier}. Meanwhile the first two conditions ensure that our initial set lies within this invariant set and that our unsafe set does not.
Notably, any invariant set containing our initial set must be a \textit{true} overapproximation of the reachable set from the initial set. This is our first contribution and is stated formally next as a corollary of Theorem~\ref{thm:BS}.

\begin{figure*}
	\centering
	\tikzset{every picture/.style={line width=0.75pt}} 

\begin{tikzpicture}[x=0.75pt,y=0.75pt,yscale=-0.8,xscale=0.8]

\draw  [fill={rgb, 255:red, 189; green, 16; blue, 224 }  ,fill opacity=0.39 ] (392.8,33.84) .. controls (392.8,25.64) and (399.44,19) .. (407.64,19) -- (485.96,19) .. controls (494.16,19) and (500.8,25.64) .. (500.8,33.84) -- (500.8,78.36) .. controls (500.8,86.56) and (494.16,93.2) .. (485.96,93.2) -- (407.64,93.2) .. controls (399.44,93.2) and (392.8,86.56) .. (392.8,78.36) -- cycle ;
\draw  [fill={rgb, 255:red, 80; green, 151; blue, 235 }  ,fill opacity=0.48 ] (258,165) .. controls (258,158.92) and (262.92,154) .. (269,154) -- (356,154) .. controls (362.08,154) and (367,158.92) .. (367,165) -- (367,198) .. controls (367,204.08) and (362.08,209) .. (356,209) -- (269,209) .. controls (262.92,209) and (258,204.08) .. (258,198) -- cycle ;
\draw  [fill={rgb, 255:red, 126; green, 211; blue, 33 }  ,fill opacity=0.34 ] (456,181.8) .. controls (456,171.42) and (464.42,163) .. (474.8,163) -- (531.2,163) .. controls (541.58,163) and (550,171.42) .. (550,181.8) -- (550,255.4) .. controls (550,265.78) and (541.58,274.2) .. (531.2,274.2) -- (474.8,274.2) .. controls (464.42,274.2) and (456,265.78) .. (456,255.4) -- cycle ;
\draw  [fill={rgb, 255:red, 80; green, 151; blue, 235 }  ,fill opacity=0.48 ] (226,29.6) .. controls (226,24.19) and (230.39,19.8) .. (235.8,19.8) -- (312.2,19.8) .. controls (317.61,19.8) and (322,24.19) .. (322,29.6) -- (322,59) .. controls (322,64.41) and (317.61,68.8) .. (312.2,68.8) -- (235.8,68.8) .. controls (230.39,68.8) and (226,64.41) .. (226,59) -- cycle ;
\draw    (501,63.86) -- (531.8,63.86) -- (542.8,63.86) ;
\draw [shift={(544.8,63.86)}, rotate = 180] [color={rgb, 255:red, 0; green, 0; blue, 0 }  ][line width=0.75]    (10.93,-3.29) .. controls (6.95,-1.4) and (3.31,-0.3) .. (0,0) .. controls (3.31,0.3) and (6.95,1.4) .. (10.93,3.29)   ;
\draw    (323.75,35.25) -- (390.8,35.25) ;
\draw [shift={(392.8,35.25)}, rotate = 180.79] [color={rgb, 255:red, 0; green, 0; blue, 0 }  ][line width=0.75]    (10.93,-3.29) .. controls (6.95,-1.4) and (3.31,-0.3) .. (0,0) .. controls (3.31,0.3) and (6.95,1.4) .. (10.93,3.29)   ;
\draw    (322,59) -- (388.8,59) ;
\draw [shift={(390.8,59)}, rotate = 180.17] [color={rgb, 255:red, 0; green, 0; blue, 0 }  ][line width=0.75]    (10.93,-3.29) .. controls (6.95,-1.4) and (3.31,-0.3) .. (0,0) .. controls (3.31,0.3) and (6.95,1.4) .. (10.93,3.29)   ;
\draw    (179.8,48.2) -- (220,48.2) ;
\draw [shift={(220,48.2)}, rotate = 181.52] [color={rgb, 255:red, 0; green, 0; blue, 0 }  ][line width=0.75]    (10.93,-3.29) .. controls (6.95,-1.4) and (3.31,-0.3) .. (0,0) .. controls (3.31,0.3) and (6.95,1.4) .. (10.93,3.29)   ;
\draw    (299.8,109.2) -- (299,146.2) ;
\draw [shift={(299.8,146.2)}, rotate = 270] [color={rgb, 255:red, 0; green, 0; blue, 0 }  ][line width=0.75]    (10.93,-3.29) .. controls (6.95,-1.4) and (3.31,-0.3) .. (0,0) .. controls (3.31,0.3) and (6.95,1.4) .. (10.93,3.29)   ;
\draw    (372.8,190.2) -- (449.33,190.2) ;
\draw [shift={(451.33,190.2)}, rotate = 179.85] [color={rgb, 255:red, 0; green, 0; blue, 0 }  ][line width=0.75]    (10.93,-3.29) .. controls (6.95,-1.4) and (3.31,-0.3) .. (0,0) .. controls (3.31,0.3) and (6.95,1.4) .. (10.93,3.29)   ;
\draw    (372,265) -- (451.33,265) ;
\draw [shift={(453.33,265)}, rotate = 180] [color={rgb, 255:red, 0; green, 0; blue, 0 }  ][line width=0.75]    (10.93,-3.29) .. controls (6.95,-1.4) and (3.31,-0.3) .. (0,0) .. controls (3.31,0.3) and (6.95,1.4) .. (10.93,3.29)   ;
\draw  [fill={rgb, 255:red, 80; green, 151; blue, 235 }  ,fill opacity=0.48 ] (257,233) .. controls (257,226.92) and (261.92,222) .. (268,222) -- (355,222) .. controls (361.08,222) and (366,226.92) .. (366,233) -- (366,266) .. controls (366,272.08) and (361.08,277) .. (355,277) -- (268,277) .. controls (261.92,277) and (257,272.08) .. (257,266) -- cycle ;
\draw    (172.8,210.2) -- (212,210.2) ;
\draw    (213,185) -- (213,235.2) ;
\draw    (213,185) -- (251.8,185) ;
\draw [shift={(253.8,185)}, rotate = 180.28] [color={rgb, 255:red, 0; green, 0; blue, 0 }  ][line width=0.75]    (10.93,-3.29) .. controls (6.95,-1.4) and (3.31,-0.3) .. (0,0) .. controls (3.31,0.3) and (6.95,1.4) .. (10.93,3.29)   ;
\draw    (211.8,235.2) -- (250.6,235.2) ;
\draw [shift={(252.6,235.2)}, rotate = 180.28] [color={rgb, 255:red, 0; green, 0; blue, 0 }  ][line width=0.75]    (10.93,-3.29) .. controls (6.95,-1.4) and (3.31,-0.3) .. (0,0) .. controls (3.31,0.3) and (6.95,1.4) .. (10.93,3.29)   ;
\draw  [fill={rgb, 255:red, 245; green, 166; blue, 35 }  ,fill opacity=0.73 ] (49,30.8) .. controls (49,25.39) and (53.39,21) .. (58.8,21) -- (168,21) .. controls (173.41,21) and (177.8,25.39) .. (177.8,30.8) -- (177.8,60.2) .. controls (177.8,65.61) and (173.41,70) .. (168,70) -- (58.8,70) .. controls (53.39,70) and (49,65.61) .. (49,60.2) -- cycle ;
\draw  [dash pattern={on 4.5pt off 4.5pt}] (40.33,28.84) .. controls (40.33,18.99) and (48.32,11) .. (58.17,11) -- (534.96,11) .. controls (544.81,11) and (552.8,18.99) .. (552.8,28.84) -- (552.8,82.36) .. controls (552.8,92.21) and (544.81,100.2) .. (534.96,100.2) -- (58.17,100.2) .. controls (48.32,100.2) and (40.33,92.21) .. (40.33,82.36) -- cycle ;
\draw  [dash pattern={on 4.5pt off 4.5pt}] (148.33,176.2) .. controls (148.33,161.18) and (160.51,149) .. (175.53,149) -- (533.13,149) .. controls (548.16,149) and (560.33,161.18) .. (560.33,176.2) -- (560.33,257.8) .. controls (560.33,272.82) and (548.16,285) .. (533.13,285) -- (175.53,285) .. controls (160.51,285) and (148.33,272.82) .. (148.33,257.8) -- cycle ;
\draw    (412.33,240) -- (450.8,240) ;
\draw [shift={(452.8,240)}, rotate = 180.28] [color={rgb, 255:red, 0; green, 0; blue, 0 }  ][line width=0.75]    (10.93,-3.29) .. controls (6.95,-1.4) and (3.31,-0.3) .. (0,0) .. controls (3.31,0.3) and (6.95,1.4) .. (10.93,3.29)   ;
\draw    (415.33,219) -- (456.8,219) ;
\draw [shift={(453.8,219)}, rotate = 180.3] [color={rgb, 255:red, 0; green, 0; blue, 0 }  ][line width=0.75]    (10.93,-3.29) .. controls (6.95,-1.4) and (3.31,-0.3) .. (0,0) .. controls (3.31,0.3) and (6.95,1.4) .. (10.93,3.29)   ;

\draw (230.8,28) node [anchor=north west][inner sep=0.75pt]   [align=left] {\begin{minipage}[lt]{53.18pt}\setlength\topsep{0pt}
\begin{center}
Simulation \\engine
\end{center}

\end{minipage}};
\draw (277.5,171.08) node [anchor=north west][inner sep=0.75pt]   [align=left] {\begin{minipage}[lt]{40.13pt}\setlength\topsep{0pt}
\begin{center}
MetaNN
\end{center}

\end{minipage}};
\draw (450,197.45) node [anchor=north west][inner sep=0.75pt]   [align=left] {\begin{minipage}[lt]{60.9pt}\setlength\topsep{0pt}
\begin{center}
Verifier\\(SMT solver)
\end{center}

\end{minipage}};
\draw (504.8,41.28) node [anchor=north west][inner sep=0.75pt]  [color={rgb, 255:red, 0; green, 0; blue, 0 }  ,opacity=1 ]  {$BC$};
\draw (325,16.4) node [anchor=north west][inner sep=0.75pt]  [color={rgb, 255:red, 208; green, 2; blue, 27 }  ,opacity=1 ]  {$\mathcal{X}_{u}$};
\draw (325.75,41.65) node [anchor=north west][inner sep=0.75pt]  [color={rgb, 255:red, 0; green, 0; blue, 0 }  ,opacity=1 ]  {$\mathcal{X}_{b}$};
\draw (418,46) node [anchor=north west][inner sep=0.75pt]   [align=left] {FOSSIL};
\draw (310,110) node [anchor=north west][inner sep=0.75pt]   [align=left] {Training \\data};
\draw (372,151.4) node [anchor=north west][inner sep=0.75pt]  [color={rgb, 255:red, 245; green, 166; blue, 35 }  ,opacity=1 ]  {$ \begin{array}{l}
BC\\
\text{cand.}
\end{array}$};
\draw (169,190.4) node [anchor=north west][inner sep=0.75pt]  [color={rgb, 255:red, 0; green, 0; blue, 0 }  ,opacity=1 ]  {$\mathcal{X}_{0}$};
\draw (266.8,231) node [anchor=north west][inner sep=0.75pt]   [align=left] {\begin{minipage}[lt]{53.18pt}\setlength\topsep{0pt}
\begin{center}
Simulation \\engine
\end{center}

\end{minipage}};
\draw (370.75,245.65) node [anchor=north west][inner sep=0.75pt]  [color={rgb, 255:red, 0; green, 0; blue, 0 }  ,opacity=1 ]  {$\mathcal{X}_{b}$};
\draw (43,27) node [anchor=north west][inner sep=0.75pt]   [align=left] {\begin{minipage}[lt]{84.38pt}\setlength\topsep{0pt}
\begin{center}
Generate training \\samples $\displaystyle \mathcal{X}_{0}$$ $
\end{center}

\end{minipage}};
\draw (54,80) node [anchor=north west][inner sep=0.75pt]   [align=left] {\bf Training};
\draw (154,260) node [anchor=north west][inner sep=0.75pt]   [align=left] {\bf Deployment};
\draw (413,221.4) node [anchor=north west][inner sep=0.75pt]  [color={rgb, 255:red, 208; green, 2; blue, 27 }  ,opacity=1 ]  {$\mathcal{X}_{u}$};
\draw (416,201.33) node [anchor=north west][inner sep=0.75pt]   [color={rgb, 255:red, 0; green, 0; blue, 0 }  ,opacity=1 ] {$\displaystyle \mathcal{X}_{0}$};

\end{tikzpicture}
	\caption{The proposed framework has two phases: (i) a training phase (top) where the training data is prepared and the MetaNN is trained; and (ii) a deployment phase (bottom) where sound reach sets are generated via neural barrier certificates. $\mathcal{X}_0=$ initial set; $\mathcal{X}_u=$ unsafe set; $\mathcal{X}_b=$ working region; $BC$= barrier certificate.
	}
	\label{alg_fig}
\end{figure*}

\begin{corollary}
\label{cor:reach}
Define $\mathcal R$ to be the set of states reachable from the initial set $\mathcal X_0$ under the dynamics \eqref{eq:system}.
Assume $B(x)$ is a barrier certificate  satisfying the conditions of Definition~\ref{def:BC}. Then, the set $\mathcal R_o = \{x\in\mathcal X\,|\, B(x)\le 0\}$ gives an over-approximation of $\mathcal R$, \ie $\mathcal R\subseteq \mathcal R_o$.
\end{corollary}

Barrier certificates can also be used to study the safety of dynamical systems when their trajectories are restricted to a subset of the state space called a \emph{working region}. This is formally stated next and will be relevant in our case study on safe autonomy.

\begin{theorem}
\label{thm:bounding}
Consider
a working region $\mathcal X_b\subseteq \mathcal X$,
the initial set $\mathcal X_0\subseteq \mathcal X_b$,
and unsafe set $\mathcal X_u\subseteq \mathcal X$.
If a function $B:\mathcal X\rightarrow\mathbb R$ satisfies the following conditions:
\begin{align}
    & B(x) \leq 0 \ \forall x  \in \mathcal X_0,
    \,\,B(x) > 0 \ \forall x  \in \mathcal X_u\cap \mathcal X_b, \nonumber\\
    & \frac{\partial B(x)}{\partial x}f(x) < 0, \ \forall x \in \mathcal X_b \ s.t. \ B(x) = 0,\label{eq:barrier4safety}
\end{align}
then the trajectories of \eqref{eq:system} evolve inside $\mathcal R_b := \{x\in\mathcal X_b\,|\, B(x)\le 0 \}$ before leaving the working region $\mathcal X_b$.
\end{theorem}
Theorem~\ref{thm:bounding} entails that we can use $\mathcal R_b$ as an over-approximation of the reach set when trajectories are restricted to $\mathcal X_b$. 

\section{Computation of Safe Reach Sets}\label{sec:method}




We utilize machine learning techniques and barrier certificate properties to efficiently compute sound reachable sets for the system modeled by \eqref{eq:system}. 
Given an initial region $\mathcal X_0$ as an \textbf{input} we seek a barrier certificate as an \textbf{output}.
By parameterizing the initial set (\eg as vertices of a box) and the barrier certificate (\eg as coefficients of a polynomial), we can train a neural network to provide barrier certificates for any general initial set. 




We present schematically our algorithm in Fig.~\ref{alg_fig}, which consists of two phases. The first stage is \emph{training}, and consists of training the MetaNN to generalize the computation of barrier certificates from initial sets. The generation of training data for this phase will be discussed shortly. 

The second phase of our approach is \emph{deployment}. Here we tap onto the MetaNN to generate reachable sets for a given initial set, by providing a candidate barrier certificate. This candidate is checked for validity using an SMT-solver. If the candidate  barrier certificate is valid, then we have successfully computed a safe reachable set. Otherwise, we can use the counter-example generated by the SMT solver to \emph{refine} the training of the MetaNN. 


\begin{remark}
The framework presented in Fig.~\ref{alg_fig} is underpinned on the continuity properties of the system trajectories with respect to its initial state. The problem of barrier certificate synthesis may be seen as a (continuous) function of this initial set $\mathcal X_0$, with similar certificates existing for similar initial sets. The MetaNN generalizes by learning over a dataset of barrier certificates. 
\end{remark}

\begin{figure*}
	\centering
	\input{figures/alg}
	\caption{Structure of online safe planning scheme. The base controller (yellow) operates in normal conditions, while the safe controller (purple) is activated to avoid unsafe regions in case of potential hazard. The MetaNN is already trained. The Verifier checks the barrier certificate (BC) candidate and calls the {\em safe controller} in case the result is unsafe. Then the process of generating a BC candidate and running the simulation engine repeats and the new candidate is checked. If the system is unsafe or the MetaNN fails to generate a valid BC within the time bound the system stops.}
	\label{case_fig}
\end{figure*}


We now discuss each part of our framework in more detail.

\subsection{Simulation Engine}
If the model dynamics are unstable the set of reachable states $\mathcal{R}$ is unbounded, which a barrier certificate cannot represent. To alleviate this, we construct a compact working region $\mathcal{X}_b$, as in Thm.~\ref{thm:bounding}.
Given an initial set $\mathcal{X}_0$ (a hyperbox), the simulation engine generates a working region for it. We simulate trajectories from the vertices of the initial set and a random sample of points inside of it. Then we take the convex hull of the generated trajectories, and take an octagonal over-approximation of the generated convex hull. This provides a working region $\mathcal{X}_b$ that corresponds to $\mathcal{X}_0$.

The working region generated by the simulation engine is key to both the \emph{training phase} and the \emph{deployment phase}.

 \noindent\textbf{Training Phase.} The working region is used to calculate the unsafe set as $\mathcal{X}_u = \mathcal{X} \setminus \mathcal{X}_b$. This unsafe set is necessary to generate barrier certificate training data using FOSSIL.
 
\noindent\textbf{Deployment Phase.} The working region is intersected with the sub-zero level set of the candidate barrier certificate to obtain the candidate reach set. Note that this means safety guarantees only hold while trajectories stay in the bounding box.



\subsection{Training of the MetaNN} 
Let $\mathcal{X}_0$ denote initial sets represented as parameterized boxes and $a \in \mathbb{R}^d$ the coefficients of a barrier certificate, $B(a, x)$, for a given template with $d$ terms. For example, if the template is a polynomial then $a$ represents the coefficients of the polynomial.
For each initial set, we compute a corresponding unsafe set $\mathcal{X}_u$ as described previously, which represents the non-reachable region of the state space. We seek to learn a mapping from $\mathcal{X}_0$ to $a$, \ie MetaNN should be able to generate barrier certificates that represent the reach sets of the system.

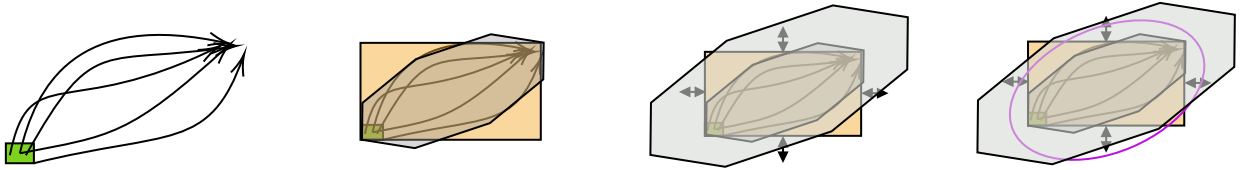
\begin{figure*}
\centering
\begin{minipage}{.23\textwidth}
\centering
	\tikzset{every picture/.style={line width=0.75pt}} 

\begin{tikzpicture}[x=0.75pt,y=0.75pt,yscale=-1,xscale=1]

\draw  [fill={rgb, 255:red, 126; green, 211; blue, 33 }  ,fill opacity=1 ] (268,63) -- (282,63) -- (282,73) -- (268,73) -- cycle ;
\draw    (275,68) .. controls (282.96,34.17) and (302.8,-4.61) .. (378.85,12.73) ;
\draw [shift={(380,13)}, rotate = 193.16] [color={rgb, 255:red, 0; green, 0; blue, 0 }  ][line width=0.75]    (10.93,-3.29) .. controls (6.95,-1.4) and (3.31,-0.3) .. (0,0) .. controls (3.31,0.3) and (6.95,1.4) .. (10.93,3.29)   ;
\draw    (282,73) .. controls (349.32,56.17) and (369.6,67.76) .. (386.49,19.48) ;
\draw [shift={(387,18)}, rotate = 108.78] [color={rgb, 255:red, 0; green, 0; blue, 0 }  ][line width=0.75]    (10.93,-3.29) .. controls (6.95,-1.4) and (3.31,-0.3) .. (0,0) .. controls (3.31,0.3) and (6.95,1.4) .. (10.93,3.29)   ;
\draw    (278,69) .. controls (312.83,6.31) and (314.98,24.81) .. (381.98,14.16) ;
\draw [shift={(383,14)}, rotate = 170.81] [color={rgb, 255:red, 0; green, 0; blue, 0 }  ][line width=0.75]    (10.93,-3.29) .. controls (6.95,-1.4) and (3.31,-0.3) .. (0,0) .. controls (3.31,0.3) and (6.95,1.4) .. (10.93,3.29)   ;
\draw    (270,69) .. controls (279.95,20.24) and (297.82,52.67) .. (373.85,14.58) ;
\draw [shift={(375,14)}, rotate = 153.14] [color={rgb, 255:red, 0; green, 0; blue, 0 }  ][line width=0.75]    (10.93,-3.29) .. controls (6.95,-1.4) and (3.31,-0.3) .. (0,0) .. controls (3.31,0.3) and (6.95,1.4) .. (10.93,3.29)   ;
\draw    (275,68) .. controls (320.77,59.04) and (328.92,59.99) .. (379.24,13.7) ;
\draw [shift={(380,13)}, rotate = 137.34] [color={rgb, 255:red, 0; green, 0; blue, 0 }  ][line width=0.75]    (10.93,-3.29) .. controls (6.95,-1.4) and (3.31,-0.3) .. (0,0) .. controls (3.31,0.3) and (6.95,1.4) .. (10.93,3.29)   ;

\end{tikzpicture}
 \end{minipage}
 \begin{minipage}{.23\textwidth}
 \centering
		\tikzset{every picture/.style={line width=0.75pt}} 

\begin{tikzpicture}[x=0.75pt,y=0.75pt,yscale=-.75,xscale=.75]

\draw  [fill={rgb, 255:red, 126; green, 211; blue, 33 }  ,fill opacity=1 ] (264,63) -- (278,63) -- (278,73) -- (264,73) -- cycle ;
\draw    (271,68) .. controls (278.96,34.17) and (298.8,-4.61) .. (374.85,12.73) ;
\draw [shift={(376,13)}, rotate = 193.16] [color={rgb, 255:red, 0; green, 0; blue, 0 }  ][line width=0.75]    (10.93,-3.29) .. controls (6.95,-1.4) and (3.31,-0.3) .. (0,0) .. controls (3.31,0.3) and (6.95,1.4) .. (10.93,3.29)   ;
\draw    (278,73) .. controls (345.32,56.17) and (365.6,67.76) .. (382.49,19.48) ;
\draw [shift={(383,18)}, rotate = 108.78] [color={rgb, 255:red, 0; green, 0; blue, 0 }  ][line width=0.75]    (10.93,-3.29) .. controls (6.95,-1.4) and (3.31,-0.3) .. (0,0) .. controls (3.31,0.3) and (6.95,1.4) .. (10.93,3.29)   ;
\draw    (274,69) .. controls (308.83,6.31) and (310.98,24.81) .. (377.98,14.16) ;
\draw [shift={(379,14)}, rotate = 170.81] [color={rgb, 255:red, 0; green, 0; blue, 0 }  ][line width=0.75]    (10.93,-3.29) .. controls (6.95,-1.4) and (3.31,-0.3) .. (0,0) .. controls (3.31,0.3) and (6.95,1.4) .. (10.93,3.29)   ;
\draw    (266,69) .. controls (275.95,20.24) and (293.82,52.67) .. (369.85,14.58) ;
\draw [shift={(371,14)}, rotate = 153.14] [color={rgb, 255:red, 0; green, 0; blue, 0 }  ][line width=0.75]    (10.93,-3.29) .. controls (6.95,-1.4) and (3.31,-0.3) .. (0,0) .. controls (3.31,0.3) and (6.95,1.4) .. (10.93,3.29)   ;
\draw    (271,68) .. controls (316.77,59.04) and (324.92,59.99) .. (375.24,13.7) ;
\draw [shift={(376,13)}, rotate = 137.34] [color={rgb, 255:red, 0; green, 0; blue, 0 }  ][line width=0.75]    (10.93,-3.29) .. controls (6.95,-1.4) and (3.31,-0.3) .. (0,0) .. controls (3.31,0.3) and (6.95,1.4) .. (10.93,3.29)   ;
\draw  [fill={rgb, 255:red, 245; green, 166; blue, 35 }  ,fill opacity=0.45 ] (263,8) -- (383,8) -- (383,73) -- (263,73) -- cycle ;
\draw  [fill={rgb, 255:red, 155; green, 155; blue, 155 }  ,fill opacity=0.39 ] (349.76,2.24) -- (384.9,7.84) -- (384.63,32.53) -- (349.1,61.84) -- (299.14,78.6) -- (264,73) -- (264.27,48.31) -- (299.8,19) -- cycle ;

\end{tikzpicture}
  \end{minipage}
   \begin{minipage}{.23\textwidth}
   \centering
	\tikzset{every picture/.style={line width=0.75pt}} 

\begin{tikzpicture}[x=0.75pt,y=0.75pt,yscale=-0.65,xscale=0.65]

\draw  [fill={rgb, 255:red, 126; green, 211; blue, 33 }  ,fill opacity=1 ] (270,100) -- (284,100) -- (284,110) -- (270,110) -- cycle ;
\draw    (277,105) .. controls (284.96,71.17) and (304.8,32.39) .. (380.85,49.73) ;
\draw [shift={(382,50)}, rotate = 193.16] [color={rgb, 255:red, 0; green, 0; blue, 0 }  ][line width=0.75]    (10.93,-3.29) .. controls (6.95,-1.4) and (3.31,-0.3) .. (0,0) .. controls (3.31,0.3) and (6.95,1.4) .. (10.93,3.29)   ;
\draw    (284,110) .. controls (351.32,93.17) and (371.6,104.76) .. (388.49,56.48) ;
\draw [shift={(389,55)}, rotate = 108.78] [color={rgb, 255:red, 0; green, 0; blue, 0 }  ][line width=0.75]    (10.93,-3.29) .. controls (6.95,-1.4) and (3.31,-0.3) .. (0,0) .. controls (3.31,0.3) and (6.95,1.4) .. (10.93,3.29)   ;
\draw    (280,106) .. controls (314.83,43.31) and (316.98,61.81) .. (383.98,51.16) ;
\draw [shift={(385,51)}, rotate = 170.81] [color={rgb, 255:red, 0; green, 0; blue, 0 }  ][line width=0.75]    (10.93,-3.29) .. controls (6.95,-1.4) and (3.31,-0.3) .. (0,0) .. controls (3.31,0.3) and (6.95,1.4) .. (10.93,3.29)   ;
\draw    (272,106) .. controls (281.95,57.24) and (299.82,89.67) .. (375.85,51.58) ;
\draw [shift={(377,51)}, rotate = 153.14] [color={rgb, 255:red, 0; green, 0; blue, 0 }  ][line width=0.75]    (10.93,-3.29) .. controls (6.95,-1.4) and (3.31,-0.3) .. (0,0) .. controls (3.31,0.3) and (6.95,1.4) .. (10.93,3.29)   ;
\draw    (277,105) .. controls (322.77,96.04) and (330.92,96.99) .. (381.24,50.7) ;
\draw [shift={(382,50)}, rotate = 137.34] [color={rgb, 255:red, 0; green, 0; blue, 0 }  ][line width=0.75]    (10.93,-3.29) .. controls (6.95,-1.4) and (3.31,-0.3) .. (0,0) .. controls (3.31,0.3) and (6.95,1.4) .. (10.93,3.29)   ;
\draw  [fill={rgb, 255:red, 245; green, 166; blue, 35 }  ,fill opacity=0.45 ] (270,45) -- (390,45) -- (390,110) -- (270,110) -- cycle ;
\draw    (330,28) -- (330,43) ;
\draw [shift={(330,46)}, rotate = 270] [fill={rgb, 255:red, 0; green, 0; blue, 0 }  ][line width=0.08]  [draw opacity=0] (8.93,-4.29) -- (0,0) -- (8.93,4.29) -- cycle    ;
\draw [shift={(330,25)}, rotate = 90] [fill={rgb, 255:red, 0; green, 0; blue, 0 }  ][line width=0.08]  [draw opacity=0] (8.93,-4.29) -- (0,0) -- (8.93,4.29) -- cycle    ;
\draw    (330,113) -- (330,128) ;
\draw [shift={(330,131)}, rotate = 270] [fill={rgb, 255:red, 0; green, 0; blue, 0 }  ][line width=0.08]  [draw opacity=0] (8.93,-4.29) -- (0,0) -- (8.93,4.29) -- cycle    ;
\draw [shift={(330,110)}, rotate = 90] [fill={rgb, 255:red, 0; green, 0; blue, 0 }  ][line width=0.08]  [draw opacity=0] (8.93,-4.29) -- (0,0) -- (8.93,4.29) -- cycle    ;
\draw    (408,77) -- (393,77) ;
\draw [shift={(390,77)}, rotate = 360] [fill={rgb, 255:red, 0; green, 0; blue, 0 }  ][line width=0.08]  [draw opacity=0] (8.93,-4.29) -- (0,0) -- (8.93,4.29) -- cycle    ;
\draw [shift={(411,77)}, rotate = 180] [fill={rgb, 255:red, 0; green, 0; blue, 0 }  ][line width=0.08]  [draw opacity=0] (8.93,-4.29) -- (0,0) -- (8.93,4.29) -- cycle    ;
\draw    (268,76) -- (253,76) ;
\draw [shift={(250,76)}, rotate = 360] [fill={rgb, 255:red, 0; green, 0; blue, 0 }  ][line width=0.08]  [draw opacity=0] (8.93,-4.29) -- (0,0) -- (8.93,4.29) -- cycle    ;
\draw [shift={(271,76)}, rotate = 180] [fill={rgb, 255:red, 0; green, 0; blue, 0 }  ][line width=0.08]  [draw opacity=0] (8.93,-4.29) -- (0,0) -- (8.93,4.29) -- cycle    ;
\draw  [fill={rgb, 255:red, 155; green, 155; blue, 155 }  ,fill opacity=0.39 ] (356.76,38.24) -- (391.9,43.84) -- (391.63,68.53) -- (356.1,97.84) -- (306.14,114.6) -- (271,109) -- (271.27,84.31) -- (306.8,55) -- cycle ;
\draw  [fill={rgb, 255:red, 215; green, 218; blue, 213 }  ,fill opacity=0.58 ] (368.39,9.06) -- (425.85,18.22) -- (425.41,58.6) -- (367.31,106.53) -- (285.61,133.94) -- (228.15,124.78) -- (228.59,84.4) -- (286.69,36.47) -- cycle ;

\end{tikzpicture}
 \end{minipage}
  \begin{minipage}{.23\textwidth}
  \centering
		\tikzset{every picture/.style={line width=0.75pt}} 

\begin{tikzpicture}[x=0.75pt,y=0.75pt,yscale=-0.65,xscale=0.65]

\draw  [fill={rgb, 255:red, 126; green, 211; blue, 33 }  ,fill opacity=1 ] (256,89) -- (270,89) -- (270,99) -- (256,99) -- cycle ;
\draw    (263,94) .. controls (270.96,60.17) and (290.8,21.39) .. (366.85,38.73) ;
\draw [shift={(368,39)}, rotate = 193.16] [color={rgb, 255:red, 0; green, 0; blue, 0 }  ][line width=0.75]    (10.93,-3.29) .. controls (6.95,-1.4) and (3.31,-0.3) .. (0,0) .. controls (3.31,0.3) and (6.95,1.4) .. (10.93,3.29)   ;
\draw    (270,99) .. controls (337.32,82.17) and (357.6,93.76) .. (374.49,45.48) ;
\draw [shift={(375,44)}, rotate = 108.78] [color={rgb, 255:red, 0; green, 0; blue, 0 }  ][line width=0.75]    (10.93,-3.29) .. controls (6.95,-1.4) and (3.31,-0.3) .. (0,0) .. controls (3.31,0.3) and (6.95,1.4) .. (10.93,3.29)   ;
\draw    (266,95) .. controls (300.83,32.31) and (302.98,50.81) .. (369.98,40.16) ;
\draw [shift={(371,40)}, rotate = 170.81] [color={rgb, 255:red, 0; green, 0; blue, 0 }  ][line width=0.75]    (10.93,-3.29) .. controls (6.95,-1.4) and (3.31,-0.3) .. (0,0) .. controls (3.31,0.3) and (6.95,1.4) .. (10.93,3.29)   ;
\draw    (258,95) .. controls (267.95,46.24) and (285.82,78.67) .. (361.85,40.58) ;
\draw [shift={(363,40)}, rotate = 153.14] [color={rgb, 255:red, 0; green, 0; blue, 0 }  ][line width=0.75]    (10.93,-3.29) .. controls (6.95,-1.4) and (3.31,-0.3) .. (0,0) .. controls (3.31,0.3) and (6.95,1.4) .. (10.93,3.29)   ;
\draw    (263,94) .. controls (308.77,85.04) and (316.92,85.99) .. (367.24,39.7) ;
\draw [shift={(368,39)}, rotate = 137.34] [color={rgb, 255:red, 0; green, 0; blue, 0 }  ][line width=0.75]    (10.93,-3.29) .. controls (6.95,-1.4) and (3.31,-0.3) .. (0,0) .. controls (3.31,0.3) and (6.95,1.4) .. (10.93,3.29)   ;
\draw  [fill={rgb, 255:red, 245; green, 166; blue, 35 }  ,fill opacity=0.45 ] (256,34) -- (376,34) -- (376,99) -- (256,99) -- cycle ;
\draw    (316,17) -- (316,32) ;
\draw [shift={(316,35)}, rotate = 270] [fill={rgb, 255:red, 0; green, 0; blue, 0 }  ][line width=0.08]  [draw opacity=0] (8.93,-4.29) -- (0,0) -- (8.93,4.29) -- cycle    ;
\draw [shift={(316,14)}, rotate = 90] [fill={rgb, 255:red, 0; green, 0; blue, 0 }  ][line width=0.08]  [draw opacity=0] (8.93,-4.29) -- (0,0) -- (8.93,4.29) -- cycle    ;
\draw    (316,102) -- (316,117) ;
\draw [shift={(316,120)}, rotate = 270] [fill={rgb, 255:red, 0; green, 0; blue, 0 }  ][line width=0.08]  [draw opacity=0] (8.93,-4.29) -- (0,0) -- (8.93,4.29) -- cycle    ;
\draw [shift={(316,99)}, rotate = 90] [fill={rgb, 255:red, 0; green, 0; blue, 0 }  ][line width=0.08]  [draw opacity=0] (8.93,-4.29) -- (0,0) -- (8.93,4.29) -- cycle    ;
\draw    (394,66) -- (379,66) ;
\draw [shift={(376,66)}, rotate = 360] [fill={rgb, 255:red, 0; green, 0; blue, 0 }  ][line width=0.08]  [draw opacity=0] (8.93,-4.29) -- (0,0) -- (8.93,4.29) -- cycle    ;
\draw [shift={(397,66)}, rotate = 180] [fill={rgb, 255:red, 0; green, 0; blue, 0 }  ][line width=0.08]  [draw opacity=0] (8.93,-4.29) -- (0,0) -- (8.93,4.29) -- cycle    ;
\draw    (254,65) -- (239,65) ;
\draw [shift={(236,65)}, rotate = 360] [fill={rgb, 255:red, 0; green, 0; blue, 0 }  ][line width=0.08]  [draw opacity=0] (8.93,-4.29) -- (0,0) -- (8.93,4.29) -- cycle    ;
\draw [shift={(257,65)}, rotate = 180] [fill={rgb, 255:red, 0; green, 0; blue, 0 }  ][line width=0.08]  [draw opacity=0] (8.93,-4.29) -- (0,0) -- (8.93,4.29) -- cycle    ;
\draw  [color={rgb, 255:red, 189; green, 16; blue, 224 }  ,draw opacity=1 ] (244.55,102.58) .. controls (234.07,78.2) and (257.89,44.54) .. (297.76,27.39) .. controls (337.63,10.25) and (378.46,16.11) .. (388.94,40.49) .. controls (399.42,64.87) and (375.6,98.53) .. (335.73,115.68) .. controls (295.86,132.82) and (255.03,126.96) .. (244.55,102.58) -- cycle ;
\draw  [fill={rgb, 255:red, 155; green, 155; blue, 155 }  ,fill opacity=0.39 ] (341.76,28.24) -- (376.9,33.84) -- (376.63,58.53) -- (341.1,87.84) -- (291.14,104.6) -- (256,99) -- (256.27,74.31) -- (291.8,45) -- cycle ;
\draw  [fill={rgb, 255:red, 215; green, 218; blue, 213 }  ,fill opacity=0.58 ] (357.39,4.06) -- (414.85,13.22) -- (414.41,53.6) -- (356.31,101.53) -- (274.61,128.94) -- (217.15,119.78) -- (217.59,79.4) -- (275.69,31.47) -- cycle ;

\end{tikzpicture}
  \end{minipage}
\caption{Four-step construction of a safe reach set $R_b$. From the left-hand side: Step 1. Running simulations from the initial set $X_0$. Step 2. Computation of the octagonal over-approximation of the simulations. Step 3. Bloating of the octagonal over-approximation. Step 4. Computation of reachable set $R_b$.}
\label{fig:stepBarrier}
\end{figure*}

\noindent\textbf{Computing safe reach sets.}
We compute (the over-approximation of) the safe  reachable region $\mathcal{R}_b$
in four steps, as illustrated in Fig.~\ref{fig:stepBarrier}.
First, we run multiple simulations from the initial set $\mathcal{X}_0$.
Second, we construct from simulation traces the working region $\mathcal{X}_b$ and octagon approximation.
Next, we bloat the octagon region by a factor $\varepsilon$, and pass its complement $\mathcal{X}_u$ to FOSSIL,
along with $\mathcal{X}_0$.
Once we obtain a valid barrier certificate from FOSSIL,
we intersect it with the working region $X_b$.
Whenever FOSSIL cannot produce a valid barrier certificate for the provided $X_b$, we further bloat the working region and call FOSSIL again.
We repeat this until a certificate is found or until a specified maximal bloating level is reached, in which case the system is deemed to be locally unsafe. 


\noindent\textbf{Training Data Generation.}
We start by generating with FOSSIL a barrier certificate $B(a, \cdot)$ for each  pair $(\mathcal{X}_0, \mathcal{X}_u)$, and by storing the coefficients $a$ of the barrier certificate, which will be our training data. Once trained, the MetaNN should map an initial set to a set of coefficients for which the barrier certificate results in a set $\mathcal{R}_b$ such that $\mathcal{R} \subseteq \mathcal{R}_b$, \ie an over-approximation of the true reach set.

The barrier certificate generation in FOSSIL uses neural network templates. FOSSIL initializes weights and biases of the neural network template randomly, which are then guided to a solution using gradient descent. 
Since the training can be done by sequentially moving the initial set gradually and iteratively through the state space, we can utilize this to find more useful training data. Specifically, we `seed' the parameters of the next initialization using the parameters of the most recent solution. In other words, we begin searching for the next barrier certificate from the previous one.
The advantage of this seeding is twofold. First, it reduces the amount of computation required to generate the training data by improving the performance of FOSSIL since we start closer to a valid certificate. Second, it allows us to take advantage of potential continuity results when generalizing over a function space.
Many barrier certificates may exist for a given initial set and template. While the resulting functions may be structurally similar, the corresponding coefficients may not be, making generalization more difficult. 
This seeding approach encourages congruous barrier certificates representations, which in turn enhances the ability of MetaNN to generalize over the function space.
If at any point we fail to find a barrier certificate for a given initial set, we simply use the most recent solution to `seed' the next attempt.

For a given data point consisting of an initial set $\mathcal{X}_0$ and a barrier certificate representation $a$, it is reasonable to assume that other valid barrier certificate representations exist near (in the Euclidean sense) to $a$. Thus, we can  generate additional data points for a fixed initial set by perturbing the representation $a$ with multiplicative noise as $\tilde{a} = a \cdot c$, where $c \sim N(1, \sigma^2)$ and $N$ is a normal distribution with unit mean and variance $\sigma^2$. 
Then, we can verify whether the \textit{new} barrier certificate constructed from $\tilde{a}$ is also a valid barrier certificate for the given initial set and corresponding unsafe (unreachable) region. If so, the generated barrier certificate can be added to the training dataset.
If $\sigma$ is sufficiently large this will provide meaningful additional data to the training procedure.

\noindent\textbf{Loss function and Network Structure.}
MetaNN  is a ReLU network trained using gradient descent with an L1 loss function, can be expressed as
\begin{equation}
	\label{eq:loss}
    \mathcal{L} = \frac{1}{n_d}\sum_{i=i}^{n_d}|a_i - \mathcal{N}(\mathcal{X}_{0_i}) |,
\end{equation}
where $|\cdot|$ denotes the absolute value, $n_d$ is the number of data points, $\mathcal{X}_0$ are input sets and $a$ is the output of the network representing the barrier certificate coefficients. 



\noindent\textbf{Online safe planning.}
Our approach can be implemented online, safe planning over bounded time horizons, as depicted in Fig.~\ref{case_fig}. 
The online phase of our approach is embedded in a loop that provides control inputs and iteratively checks for system safety. The loop uses two controllers: a `base' controller for normal use and a `safe' controller which takes over in emergency situations. 
The base controller provides inputs to control the system over a short time horizon; such inputs and the current system state are used to compute a candidate reach set. 
This is passed to the SMT solver, which decides whether the system is safe (for the current time horizon), \ie whether the computed reach set and the unsafe set do not intersect. If the SMT solver cannot guarantee an empty intersection, then the safe controller is activated and will provide input controls for the current time horizon. 
However, if the verifier cannot guarantee the system is safe even for the input from the safe controller, then the system stops. Otherwise, the inputs provided by the base controller are applied to the actual system, and the loop repeats for the next time horizon. 
In the current implementation, to generate an initial set for the next time horizon we use a center trajectory as the dynamics of the system. Any trajectory can be used instead and it does not affect the soundness of the algorithm. In addition, we bloat around the last state of the trajectory to account for sensing errors of real-world models.

\section{Simulation Results}\label{sec:experiments}
In this section, we apply our framework for the computation of safe reach sets on a number of linear and nonlinear autonomous models. Then, we present the application in an autonomous driving setup.
We have implemented our algorithms in Julia, using the LazySets package of JuliaReach~\citep{bogomolov2019juliareach} for set operations and the DifferentialEquations.jl package 
to perform simulations, the PyCall.jl\footnote{\url{https://github.com/JuliaPy/PyCall.jl}} package to call FOSSIL and dReal. We utilize PyTorch with CUDAt to train the MetaNN.

For all case studies, we use a second-order polynomial as a template for the barrier certificate.


\noindent\textbf{Validation for linear systems.}
We performed the data generation and MetaNN training steps described in the previous section for two linear systems.

Next, we validated the reliability of our approach at generating safe reach sets for a separate \textit{test} dataset. The test dataset is composed of input-output pairs $(\mathcal{X}_0, a)$, where $\mathcal{X}_0$ are input sets and $a$ are the coefficients of a barrier certificate for a candidate reach set. The validation consists in certifying whether the set $\mathcal{R}_b = \{ x \in \mathcal{X}_b | B(x) \leq 0\}$ constructed using the candidate certificate computed by MetaNN is in fact a correct over-approximation of the true reach set. This is equivalent to checking whether the candidate barrier certificate is valid within the working region, for the given initial set and an empty unsafe set. This can be achieved by asking an SMT-solver to seek a counterexample for the barrier certificate's validity. If no counterexample is found, this is a successful output of the MetaNN. Our validation consists of finding the percentage of successful outputs obtained over the test inputs.




\subsubsection{Linear system with real eigenvalues} 
We first consider a simple linear system with real eigenvalues:
\begin{equation}
\label{eq:lin-real}
\begin{cases}
\frac{d}{dt}x(t) = -x, & x(0) = x_0\\
\frac{d}{dt}y(t) = -2\cdot y, & y(0) = y_0.\\
\end{cases}
\end{equation}
An example barrier certificate generated for this model is shown in Fig.~\ref{fig:simpleStable}. 
We train and test the MetaNN on different initial sets of different size, but within the same specified domain. The test data consist of 1000 randomly sampled initial sets of random size. The proportion of successful outputs over the test dataset is 99.3\%. 

\subsubsection{Linear system with complex eigenvalues.} 
\begin{figure}
	\centering
\begin{minipage}{0.49\textwidth}
  \centering
	\begin{minipage}{.65\textwidth}
	  \centering
	\includegraphics[width=1.\textwidth]{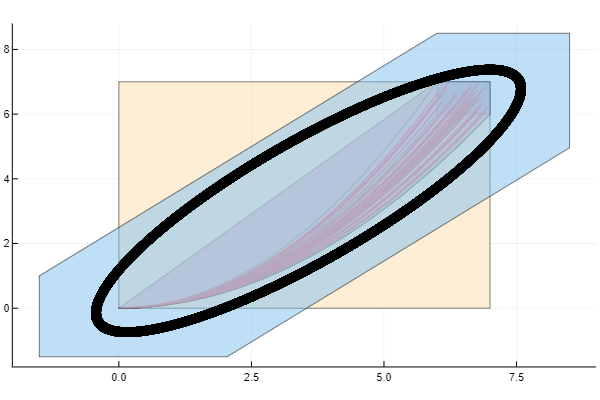}
	\subcaption[SimpleStable]{Linear system with real eigenvalues.}
	\label{fig:simpleStable}
	\end{minipage}
\end{minipage}%

\begin{minipage}{0.49\textwidth}
  \centering
	\begin{minipage}{.65\textwidth}
	  \centering
	\includegraphics[width=1.\textwidth]{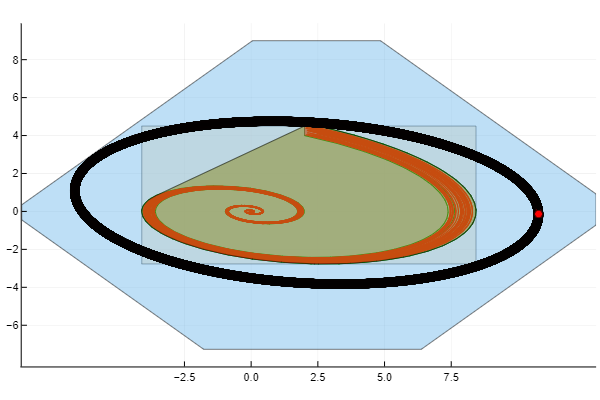}
	\subcaption[SpiralStable]{Linear system with complex eigenvalues.}
	\label{fig:spiral}
	\end{minipage}
\end{minipage}
%
\caption{Safe reach sets for the linear systems with real \eqref{eq:lin-real} and complex \eqref{eq:lin-complex} eigenvalues. The red traces are simulation trajectories. The inner and outer octagons are, respectively, the convex hull of the generated trajectories, and its over-approximation. The yellow rectangle is a bounding box. The black ellipsoid is the zero level-set of the validated barrier certificate.}
\end{figure}

Next, we consider a linear system with complex eigenvalues:
\begin{equation}
\label{eq:lin-complex}
\begin{cases}
\frac{d}{dt}x(t) = y, & x(0) = x_0\\
\frac{d}{dt}y(t) = -0.2\cdot x -0.2\cdot y, & y(0) = y_0.\\
\end{cases}
\end{equation}
A sample barrier certificate generated for this model is shown in Fig.~\ref{fig:spiral}. Similarly to the previous model, we have trained a MetaNN, and the correctness of the obtained neural barrier certificates is 99.2\% over 1000 randomly sampled initial sets of random size.

\smallskip

\noindent\textbf{Online safe path planning for nonlinear vehicle dynamics.}
Consider
\begin{equation}
\label{eq:car}
\begin{cases}
\frac{d}{dt}x(t) = v \cos \theta(t), & x(0) = x_0\\
\frac{d}{dt}y(t) = v \sin \theta(t), & y(0) = y_0\\
\frac{d}{dt}\theta(t) = \omega, & \theta(0) = \theta_0,\\
\end{cases}
\end{equation}
where $(x,y)$ indicate the location in two-dimensional space and $\theta$ is an angle describing the orientation of the car. We assume that the velocity $v$ and the rotational speed $\omega$ are inputs that are selected by a feedback controller.

We seek to design $v$ and $\omega$ such that the car, starting from an initial region, will reach a target region $\mathcal X_r$, while avoiding obstacles that are present over the $(x,y)$ domain. Here, we consider obstacles as static boxes, however the online implementation of our approach can in principle also handle dynamically-changing obstacles.
We emphasize that the control inputs ought to be executed in a provably safe manner, with sound certificates.

The literature in path planning is mature but it generally neglects the dynamics of the vehicle, and most often does not come with formal safety guarantees. We show that we can first neglect the obstacles and design a baseline controller for steering the dynamics from the initial set to the target region $\mathcal X_r$. We then show how the scheme presented in Sec.~\ref{sec:method} can be used to provide sound safety guarantees on the executed control inputs.

We have designed the following baseline controller for reaching the target:
\begin{equation*}
\begin{cases}
v(t) = \alpha_1\left(1-\exp(-\alpha_2\|(x(t)-x_r,y(t)-y_r)\|_2)\right)\\
\omega(t) = \alpha_3(\theta_r(t)-\theta(t))\\
\theta_r(t):=\arctan\left(\frac{y(t)-y_r}{x(t)-x_r}\right),
\end{cases}
\end{equation*}
where $(x_r,y_r)\in\mathcal X_r$ is a reference point within the target region and $\theta_r(t)$ is the relative angle between current and target location. The angular speed $\omega(t)$ is designed to steer the car from the current orientation towards the reference angle. The speed is  also proportional to the distance from the reference point passed through an exponentially decaying function, thus slowing down the vehicle when approaching the reference point.
Note that this simple controller can be replaced with any path planning algorithm that can be synthesized over \eqref{eq:car}.

The above baseline controller does not consider obstacles and does not come with any safety guarantees. These desirable features can be attained with our safe reach set approach. First, let us note that the dynamics in \eqref{eq:car} admit two structural properties that are useful in the computation of barrier certificates. While these properties are {\em not} specifically required by our approach, they can be leveraged to reduce the computational burden.

\smallskip

\noindent\textbf{Property 1.} The dynamics are shift invariant with respect to location $(x,y)$ for fixed control inputs $(v,\omega)$: if $(x(t),y(t),\theta(t))$ is a solution from an initial condition $(x_0,y_0,\theta_0)$, then $(x(t)-x',y(t)-y',\theta(t))$ is a solution from the initial condition $(x_0-x',y_0-y',\theta_0)$. This means the computation of barrier certificate can always be done for initial sets that are centered at the origin $(0,0)$.

\smallskip

\noindent\textbf{Property 2.} The dynamics are symmetric with respect to the $x$-axis. If $(x(t),y(t),\theta(t))$ is a solution from an initial condition $(x_0,y_0,\theta_0)$ with fixed control inputs $(v,\omega)$, then $(x(t),-y(t),-\theta(t))$ is a solution from the initial condition $(x_0,-y_0,-\theta_0)$ with fixed control inputs $(v,-\omega)$. This means we can restrict the computations to $\omega\ge 0$.

\smallskip




\begin{figure*}[ht]
	\centering
\begin{minipage}{0.45\textwidth}
  \centering
\includegraphics[width=1.\textwidth]{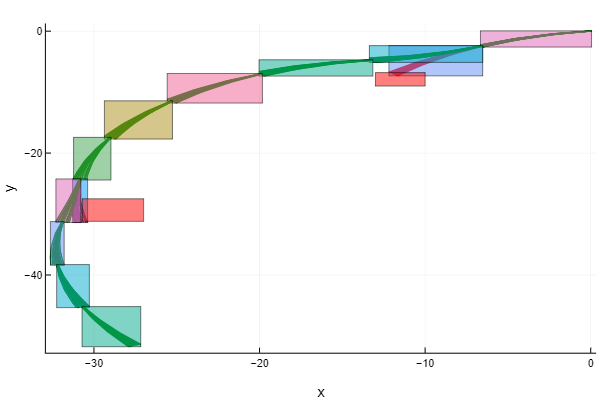}
\subcaption[ReachSet]{Reachable sets for a total time horizon $T=20$ along with unsafe occurrences at step 2 and step 7; the car starts in the top-right corner.}\label{fig:case1a}
\end{minipage}
\hspace{2ex}
\begin{minipage}{0.4\textwidth}
  \centering
	\begin{minipage}{0.7\textwidth}
	  \centering
	\includegraphics[width=1.\textwidth]{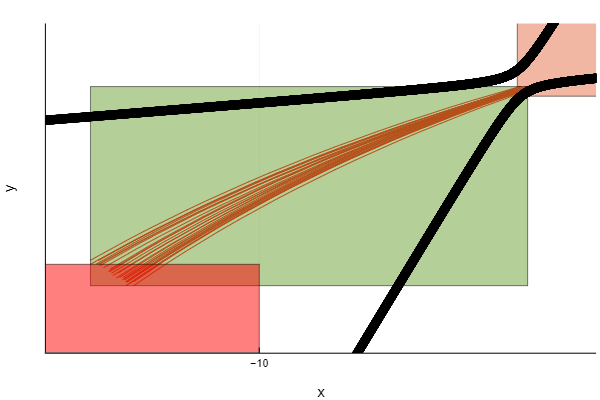}
	\end{minipage}
	\begin{minipage}{0.7\textwidth}
	  \centering
	\includegraphics[width=1.\textwidth]{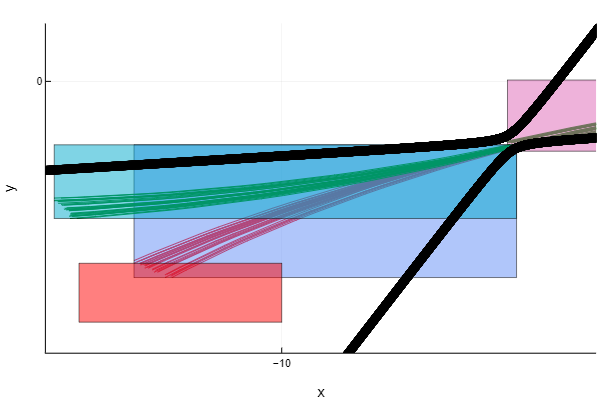}
	\end{minipage}
	\subcaption[Step2]{Original unsafe control (top) along with the safe control (bottom) for step 2.}\label{fig:case1b}
\end{minipage}%
\caption{Plot of the trajectory of the autonomous driving case study for $T=20$. Each block corresponds to $T_r=2$. Red traces correspond to trajectories which lead to unsafe region. Green trajectories are safe. Red sets represent unsafe regions. Black curves are zero-level set of barrier certificates.}
\end{figure*}

We choose the values $\alpha_1 \!=\! 0.2$, $\alpha_2 \!=\! 3.54$, and $\alpha_3 \!=\! 0.06$ for the baseline controller, and train MetaNN over the domain $v \in [1.5;10], \ \omega \in [0;0.125]$ and $\theta \in [0;6.5]$ by generating 5414 valid barrier certificates. The trained MetaNN gives a 99.5\% correctness over 1000 randomly sampled initial sets of random size. It takes 20 minutes to train the model with CUDA (NVIDIA GeForce RTX 3060 Laptop GPU) for 750000 epochs.
%
%
Fig.~\ref{fig:case1a} shows the trajectories of the car over a time horizon $T= 20$ in which safe reachability is computed every $T_r = 2$, thus corresponding to ten iterations of the procedure
in Fig.~\ref{case_fig}.


\noindent\textbf{Online Certification of Safe Reach Sets.}
%
We now study online correctness guarantees for the safe reach sets constructed by the procedure.
We wish to provide a guarantee that for a chosen control action, the vehicle does not collide with an obstacle over some fixed horizon. Note that, as per Theorem \ref{thm:bounding}, any such guarantee is conditional on the vehicle remaining within the working region $\mathcal{X}_b$: if the vehicle leaves this region within the time horizon our approach can no longer guarantee safety.

This certification check involves asserting that the function constructed from the output of  MetaNN, which we denote as $B(x)$, is a valid barrier certificate. Define the unsafe set $\mathcal{X}_u$ as the union of all box-shaped obstacles. The initial set $\mathcal{X}_0$ is a box containing the vehicle's true location. The size of this initial set accounts for the uncertainty in the location of obstacles, and to take care of the change in location of the vehicle that occurs during the computation time allocated to the generation of the safe reach sets: this vouches for the online implementation of our scheme. Finally, the working region $\mathcal{X}_b$ is computed as in Sec.~\ref{bc4rs}. Due to the trigonometric terms present in the system dynamics \eqref{eq:car}, dReal is a suitable choice of SMT solver.

We report in Figs.~\ref{fig:case1b} an example of the control action that would result in unsafe behavior, and its resolution. For cases when the dynamics leads to the unsafe region we use the same controller to avoid obstacles. We exploit the unsafe state as the target region and take a complement of the output angular speed to steer away. We note that each iteration of the algorithm (see Fig.~\ref{case_fig}) takes about $0.03$ seconds to execute on a standard laptop for $T_r=2$.

If dReal is unable to find a counter-example, the reach set is certified as being a safe over-approximation of the true reach set, and can safely proceed with the control action over the time horizon. Meanwhile, the existence of a counter-example does not imply that our control action is inherently dangerous, but rather that we cannot assert that it is safe. Our immediate response, as depicted in Fig.~\ref{fig:case1b}, is to select a different control input to attempt to steer the car safely. 

\begin{figure}
\centering
	\includegraphics[width=0.35\textwidth]{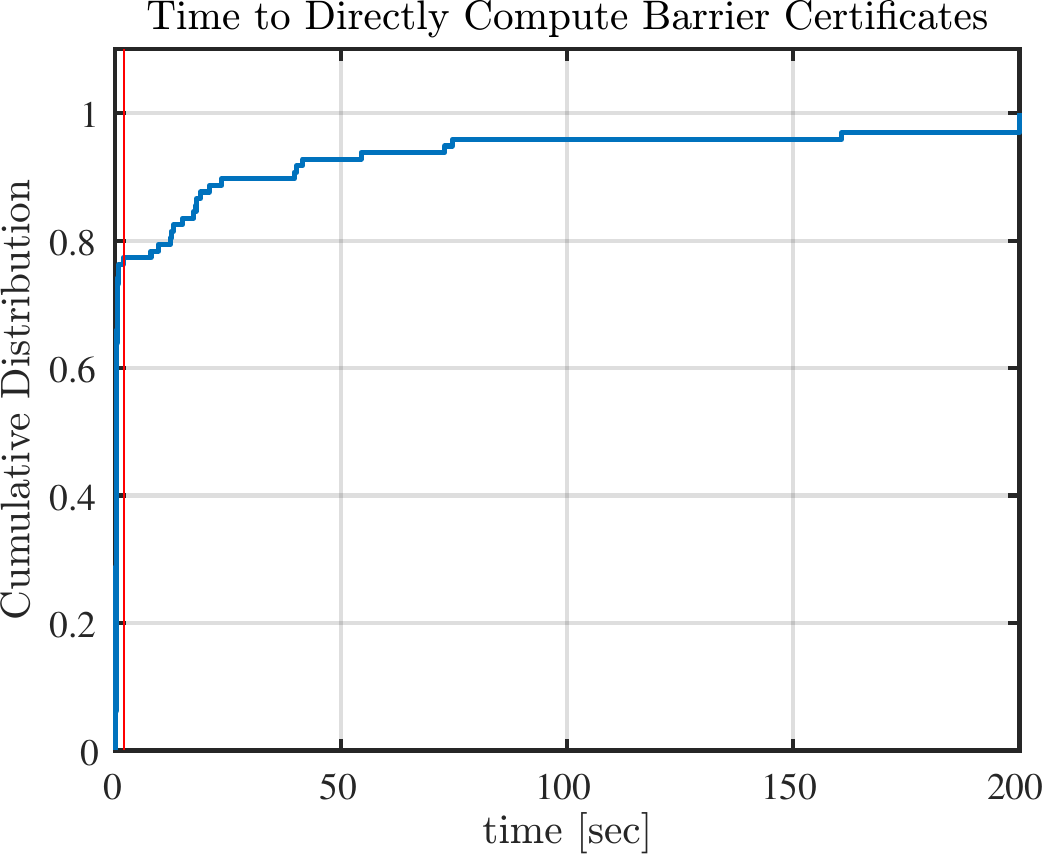}
  \caption{Cumulative distribution of time taken to generate barrier certificates via FOSSIL instead of MetaNN. The red vertical line is the acceptable time bound for the car model (2 seconds). FOSSIL synthesizes only $78\%$ valid barrier certificates by FOSSIL in this time.}
  \label{fig:cdf}
\end{figure}

\noindent\textbf{Comparison with Direct Online Certification of Safe Reach Sets.}
As demonstrated with the car model case study, our proposed MetaNN enables us to perform controller synthesis together with online certification. In order to compare the performance of the MetaNN {\em vs.}~FOSSIL, we take the same controller synthesized by the MetaNN and give it directly to FOSSIL for online certification. Figure~\ref{fig:cdf} demonstrates the cumulative distribution of the computational time taken to synthesize 100 barrier certificates directly by FOSSIL. Note, that for the car model under consideration we utilize $T_r=2$, which means if the barrier certificate is not computed within 2 seconds, online certification is not possible and an auxiliary controller should be employed. We report that only $78\%$ of instances are verified by FOSSIL on time. In addition, three cases reach time-out (200 seconds), which means that FOSSIL cannot generate certified barrier certificates with the provided specifications of the model. In Table~\ref{tbl:stat_fossil} we report the statistics of the computational time for the certification process directly from FOSSIL. We can observe that FOSSIL demonstrates divergence in computational time for different instances of the model, \eg{maximum computation time for one instance is 160s, while the minimum is 0.11s and the average is 7.11s.}

\begin{table}
\centering
\arrayrulecolor{black}
\ADLnullwidehline
\begin{tabular}{l!{\color{black}\vrule}r!{\color{black}\vrule}r}
    & All (s)     & Valid (s)    \\ 
\hline
Total & 1290.43 & 690.43  \\ 
Avg & 12.90 & 7.11  \\ 
min & 0.11 & 0.11  \\ 
max & 200      & 160.63 \\
\end{tabular}
\arrayrulecolor{black}
 \caption{Statistics of the computational time to synthesize valid barrier certificates for 100 instances on the car model. FOSSIL was not able to find a barrier certificate in 3 instances within the fixed time bound (200s). These exceptions are excluded in the Valid column.}
 \label{tbl:stat_fossil}
\end{table}

The computational time for our MetaNN remains instead constant for each instance. It takes less than 0.7s to extract a barrier certificate from the neural network and verify it with dReal.  In addition, we were able to obtain valid barrier certificates for the instances where FOSSIL timed out.  Therefore, by using the MetaNN and offline training, we not only extend the capabilities of barrier certificate verification to synthesis of parameterized controllers, but also
increases the success rate of online barrier certificate generation within two seconds from 78\% to 99\%. This is at the cost of offline training that takes 10 hours in our case study.

 \vspace{-1ex}
\section{Discussion and Conclusions}\label{sec:conclusions}
 \vspace{-2ex}
In this paper, we developed a novel approach for sound computation of reach sets of dynamical systems using level sets of barrier certificates. We developed a meta-neural network that is trained offline on samples of barrier certificates and is used online, integrated with an SMT solver, for ensuring safety.
%
Our framework can take any technique for computing barrier certificates and use it for generating training data. As a result, the offline training step of our approach highly depends on the scalability and precision of such a computational technique. In the future, we plan to explore alternative computational techniques as the core of our framework and compare against newly developed online reach set computational tools.


%
%

\begin{ack}
 The research of S.~Soudjani was supported by the following grants: EPSRC EP/V$043676$/$1$, EIC $101070802$, and ERC $101089047$.
\end{ack}




                                                   
\end{document}